\documentclass{article}

\usepackage{authblk}
\usepackage{graphicx}
\usepackage{dcolumn}
\usepackage{bm}
\usepackage{amsmath,amssymb,amsthm}
\usepackage{xspace}
\usepackage{cleveref}
\usepackage{booktabs}
\usepackage{fullpage}

\usepackage{xfrac}
\usepackage{siunitx}
\usepackage{acronym}
\usepackage{graphicx}
\usepackage{dcolumn}
\usepackage{bm}
\usepackage{amsmath,amssymb,amsthm}

\graphicspath{{./images/}}

\newacro{DPD}{dissipative particle dynamics}
\newcommand{\DPD}{\ac{DPD}\xspace}
\newacro{GPR}{gaussian process regression}
\newcommand{\GPR}{\ac{GPR}\xspace}

\renewcommand{\vec}[1]{\bm{#1}}
\newcommand{\ymr}{Mirheo\xspace}
\newcommand{\Rey}{\ifmmode \mathrm{Re}\else Re\fi}
\newcommand{\lb}{\left(}
\newcommand{\rb}{\right)}
\newcommand{\sublabel}[1]{\textbf{#1}}
\newcommand{\SM}{Supplemental Material}

\author[1]{Dmitry Alexeev}
\author[2]{Sergey Litvinov}
\author[3]{Athena Economides}
\author[2]{Lucas Amoudruz}
\author[4]{Mehmet Toner}
\author[2]{Petros Koumoutsakos}
\affil[1]{NVIDIA, Zurich, Switzerland}
\affil[2]{Computational Science and Engineering Laboratory, Harvard University, Cambridge, MA 02138, USA}
\affil[3]{Institute of Neuropathology, University Hospital Zurich, University of Zurich, Zurich, Switzerland}
\affil[4]{Center for Engineering in Medicine and Surgical Services, Massachusetts General Hospital, Harvard Medical School, Boston, MA 02114, USA}

\begin{document}
\title{Inertial Focusing of Spherical Particles: \\ The Effects of Rotational Motion}
\date{\today}

\maketitle

\begin{abstract}
  The identification of cells and particles based on their transport properties in microfluidic devices is crucial for numerous applications in biology and medicine.
  Neutrally buoyant particles transported in microfluidic channels, migrate laterally towards stable locations due to inertial effects.
  However, the effect of the particle and flow properties on these focusing positions remain largely unknown.
  We conduct large scale simulations with dissipative particle dynamics, demonstrating that freely moving particles exhibit significant differences in their focusing patterns from particles that are prevented from rotation.
  In circular pipes, we observe drastic changes in rotating versus non-rotating focusing positions.
  We demonstrate that rotation-induced lateral lift force is significant, unlike previously believed, and is linearly dependent on the rotation magnitude.
  A simple phenomenological explanation extending existing theories is presented, that agrees well with our numerical findings.
  In square ducts, we report four face-centered stable positions for rotating particles, in accordance with experimental studies on a range of Reynolds numbers $50 \leq \Rey \leq 200$.
  However, non-rotating particles stay scattered on a concentric one-dimensional annulus, revealing qualitatively different behavior with respect to the free ones.
  Our findings suggest new designs for micro-particle and cell sorting in inertia-based microfluidics devices.
\end{abstract}

\section{Introduction}

Inertial microfluidics is opening up numerous possibilities for fast and clogging-free focusing and sorting of biological cells and microparticles~\cite{DiCarlo2009a,Zhang2016,berlanda2020recent,peng2024sheathless}.
In inertial transport, particles migrate across streamlines and stabilize in certain lateral locations in the flow field.
Inertia is essential in this cross-streamline movement, as lateral translation is hindered in Stokes flow \cite{stone2004engineering,martel2014}.

In contrast to traditional cell sorting techniques~\cite{martel2014}, such as filtering~\cite{McFaul2012} or exploitation of lateral displacement~\cite{Karabacak2014}, inertial manipulation is free from externally applied fields or direct mechanical interaction with obstacles.
The first studies on inertial migration were reported in the work of Segr{\'{e}} and Silberberg~\cite{Segre1962}.
These studies reported that small particles submerged in the circular pipe flow concentrate at the annulus of radius ${\sim} 0.6 R$.
More recently, other studies~\cite{Bhagat2008,DiCarlo2009,Hur2011a,Amini2014a} have shown that these migrations occur in channels of any cross-section shape and at a wide range of Reynolds numbers (from 10 to 1000), for rigid particles, liquid droplets and cells.

Interestingly, the lateral positions where the particles focus vary significantly with Reynolds numbers, channel geometry and particle size.
Matas et al.~\cite{Matas2004a,Matas2004b} showed experimentally that the equilibrium particle position in circular pipe shifts outward as the channel Reynolds number increases.
Choi et al.~\cite{Choi2010} confirmed that bigger particles stabilize closer to the center of the pipe.
The same behavior was reported for square channels~\cite{DiCarlo2009}, with, however, an important difference: only four stable lateral positions are observed at intermediate Reynolds numbers~\cite{Miura2014}, in contrast to the stable annulus in circular pipes.
Migration to those points happens in two stages: during the first, faster, stage the particles gather into a deformed ring, similar to the original Segr{\'{e}}-Silberberg annulus.
During the second, slower, stage the particles move towards the four points next to the centers of the channel faces.
This focusing pattern makes it possible to use the square channels to precisely collect microscale particles and cells at the channel outlet and allows for fast and non-invasive methods for sorting and filtering~\cite{DiCarlo2009}.

As the small inertial effects are difficult to study analytically, various restrictions and assumptions on the problem setup were made in the theoretical works.
In particular, the particle size was assumed to be much smaller than the diameter of the channel~\cite{Saffman1968,Asmolov1999}.
However, the experimental verification of these findings is complicated, as small particles show very slow lateral migration.
Recent works~\cite{Hood2015,Hood2016} present an asymptotic theory for larger particles in square channels.
Their analysis captures experimental trends; however,~\cite{Nakagawa2015} displays notable inaccuracies in the near-wall region and is still limited to relatively small particles: ${\sim} 0.15$ of the channel size.
Overall, there is a consensus among the theoretical works that the influence of particle rotation on the lift force, and thus on the
focusing position, is negligible.
Numerical studies alleviate the restriction on the particle size, shape and the channel cross-section~\cite{shi2023numerical}.
Feng~\cite{Feng1994} used two-dimensional finite element simulations to assess the forces acting on a rigid particle in Couette and Poiseuille flows, confirming that there exists a certain stable lateral position.
They identified three main components of the lift force: wall force, force due to the non-uniform shear rate, and force associated with the particle rotation (Magnus effect), noting once again that the rotation-attributed force is much smaller than the rest.
Chun and Ladd~\cite{Chun2006} performed a three-dimensional study of the particle focusing positions in the square duct.
Although they observed the two-stage migration process, their final stable points were both corners and face centers even for low Reynolds around 100, which contradicts experiments.
A good agreement to the experimental data~\cite{Miura2014} was achieved by Nakagawa~\cite{Nakagawa2015} with the immersed boundary method.
They demonstrated that up to the critical channel Reynolds number of about $260$, only the face centers are stable, and after that corners become stable too.
Liu~\cite{Liu2015} carried out experimental and numerical study of the particle behavior in the channel with rectangular cross-section and observed similar focusing patterns as for the square channel.
They also presented results for different particle sizes, confirming that bigger particles tend to stay closer to the channel center.
Harding and Stokes~\cite{harding2023inertial} studied the inertial focusing of neutrally buoyant spherical particles in curved microfluidic ducts at moderate Dean numbers, using regular perturbation expansions.
They found that variations in the Dean number cause a change in the axial velocity profile of the background flow which influences the inertial lift force on a particle.
More recently, the authors also studied the effects of trapezoidal ducts~\cite{harding2024inertial}.
In~\cite{hafemann2023inertial} simulations of particulate flows in square ducts with oblate and prolate particles revealed that inertial migration causes particles to focus in specific cross-sectional regions, with non-spherical particles occupying different positions than spherical ones.
Investigations in a wavy channel~\cite{mao2023particle} demonstrated that interactions between the zeroth-order lift force and the particle-free flow largely determine the focusing locations.

In the present work we employ the method of \DPD{}~\cite{hoogerbrugge1992,pivkin2010dissipative,espanol2017perspective,owen2022numerical} to study the inertial focusing of rigid particles in the circular and square channel flows.
Our results agree well with experimental data and previous numerical studies.
In contrast to the recent study by Huang et al.~\cite{Huang2018}, we observe a broad range of \Rey{} and particle sizes where \DPD{} is applicable (see \SM{}~\cite{SM}, which also contains references~\cite{Espanol1995,Groot1997,Fan2006,rozmanov2010robust,Revenga1998,Fedosov2008}).

We investigate the influence of particle rotation on migration patterns within a circular pipe. Our study examines a range of particle sizes and channel Reynolds numbers. Contrary to the commonly held assumption that rotation-induced forces are negligible, we observe significant differences in the migration behavior of freely rotating and non-rotating particles, consistent with experimental findings~\cite{Oliver1962}. Our results reveal that the rotation-induced force consistently acts toward the pipe wall, aligning with the force generated by the non-uniform shear rate.
Starting with the latter force explanation by Feng~\cite{Feng1994}, we take into account flow distortion by the particle to arrive at a similar explanation for the order of magnitude and direction of the rotation-induced force.
We also provide a scaling law which describes the force due to rotation in terms of the channel Reynolds number, particle size, and angular velocity.
Linear dependence of this force on the rotation agrees with our explanation.

In the square duct we demonstrate the mid-face equilibrium positions of the particles are reached in the previously reported two-stage process.
We find that the particles with inhibited rotation show no secondary migration stage, staying scattered at the concentric annulus.
We also show that unlike hypothesised by Zhou~\cite{Zhou2013}, the secondary migration is not solely governed by the particle rotation, but rather by more complex interplay of the rotation and wall force.
Our findings unveil possibilities for novel designs of microfluidic filtering devices, allowing separation of objects based on the size and rotation.

\section{Methods}

We consider straight pipes of circular and square cross sections filled with a Newtonian, incompressible fluid, with periodic boundary conditions to mimic long pipes.
We model the suspended particle as a spherical, non-deformable object with the same density as the fluid.
To simulate this system, we employ the \DPD{} method within the software \ymr{}~\cite{alexeev2020mirheo}, which has been extensively validated for incompressible flows with suspended biological cells~\cite{amoudruz2023stress,amoudruz2024volume}, and suspended rigid objects~\cite{alexeev2020mirheo,amoudruz2022simulations}.
In addition, several cases of particles in circular pipes and square ducts are validated against experimental work~\cite{Choi2010,Miura2014} and numerical work~\cite{DiCarlo2009,Liu2015,Nakagawa2015} in the \SM{}~\cite{SM}.
The parameters of the numerical model that were used in this work are reported in the \SM{}~\cite{SM}.

\label{sec:setup}
\begin{figure}
    \centering
    \includegraphics[width=0.9\columnwidth]{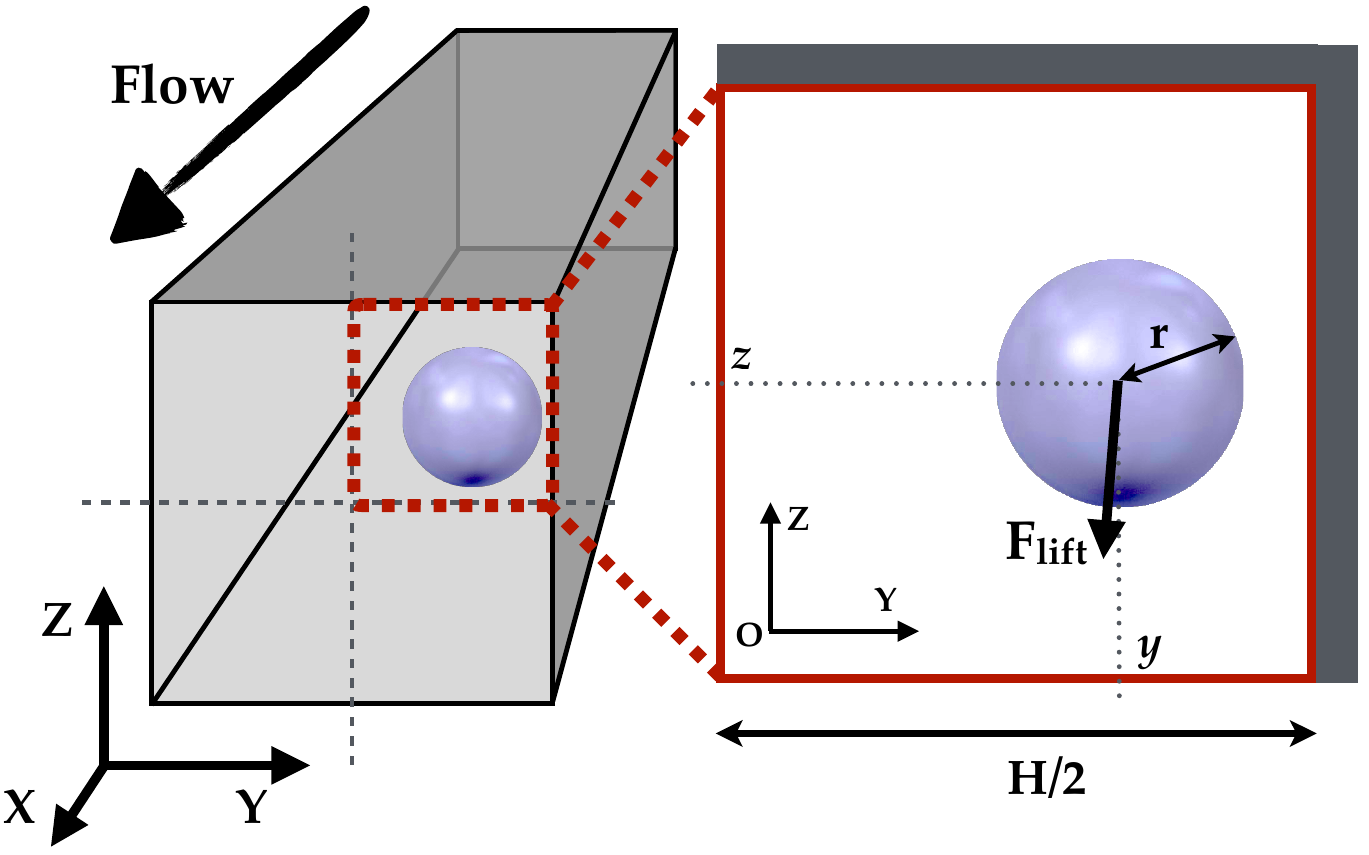}
        \caption{The particle is translated with the flow along $Ox$
          and experiences lateral forces along $Oy$ and $Oz$.  Its
          position is characterized by $y$ and $z$ with origin at the
          pipe center.}
    \label{fig:inertial_setup}
\end{figure}

The simulation describes the neutrally buoyant spherical particles of radius $r$ moving in the channel flow with translational velocity $\vec{V}$ and rotational velocity $\vec{\Omega}$ (\Cref{fig:inertial_setup}). We define $\mathbf{F}_{particle} = \mathbf{F}_{f} + \mathbf{F}_{ext}$ as the total force acting on the particle, including the hydrodynamic force and externally applied force;
$\mathbf{T}_{particle} = \mathbf{T}_{f} + \mathbf{T}_{ext}$ -- the total torque acting on the particle, including the hydrodynamic torque and externally applied torque.
Moreover, $\vec{u}$ denotes the fluid velocity, $p$ -- pressure, $\rho$ -- density, and $\nu = \sfrac{\eta}{\rho}$ -- kinematic viscosity.
Finally, we assume no-slip boundary conditions on the channel walls and the particles.

\begin{figure}
  \centering
    \includegraphics[width=0.9\linewidth]{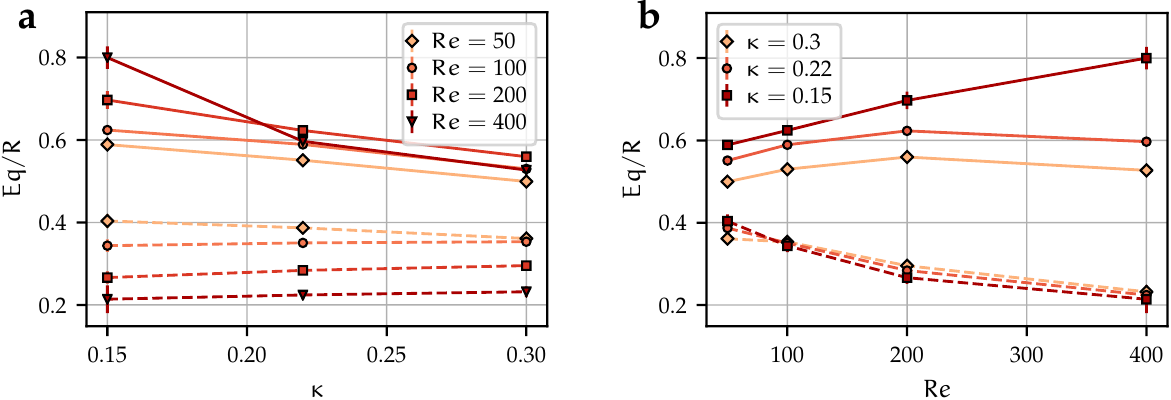}
        \caption{ Equilibrium positions $Eq$ of freely rotating (solid
          lines) and non-rotating (dashed lines) particles against
          particle-to-channel size ratio $\kappa$ (\sublabel{a}) and
          against channel Reynolds number \Rey{}\ (\sublabel{b}) in a
          circular pipe. }
        \label{fig:inertial_rigid_tube_equilibria}
\end{figure}

We denote the square channel side length as $H$ and for uniformity the circular pipe radius as $R = H/2$.
We also introduce the particle-to-channel size ratio as $\kappa = 2r/H = r/R$.
The flow is driven along the channel axis by the pressure gradient $\nabla p_{ext}$.
We define the channel Reynolds number as $\Rey{} = U_{avg}\rho H /\eta$, where $U_{avg}$ is the average velocity of the undisturbed flow,
\begin{equation}
        U_{avg}^{circle} = \frac{H^2 \nabla p_{ext}}{32 \eta}, \quad
        U_{avg}^{square} = \frac{H^2 \nabla p_{ext}}{28.45415 \eta}.
\end{equation}
The non-dimensional lift coefficient for a sphere experiencing lateral force $F_l$ is given by:
\begin{equation}
        C_l = \frac{F_l H^2}{\rho U_{avg}^2 (2r)^4}.
\end{equation}
Furthermore we define the particle Reynolds number as $\Rey_p = \kappa^2 \Rey$~\cite{Feng1994}.
To compute the forces acting on the particle at a given lateral position, we restrict the particle motion along $Oy$ and $Oz$ axes, while allowing it to freely translate with the flow along $Ox$.
The instantaneous forces acting on the particle along $Oy$ and $Oz$ are noisy due to the stochastic nature of the \DPD method.
Therefore, we average these forces over time after the simulation has equilibrated.
The simulation is considered at equilibrium after the velocity of the particle along $x$ reached a constant value.
We use periodic boundary condition in $x$ and keep the length of the channel at least $15$ times larger than the particle diameter~\cite{Nakagawa2015}.

We interpolate the averaged forces with
\GPR~\cite{Rasmussen2004}. This technique allows us to take the
uncertainty in the mean estimator and provide error estimates in the
interpolation.

\section{Results and discussion}
\subsection{Circular pipe}
We first study the inertial migration of particles in the circular
pipe.  We compute the lateral lift forces for multiple parameter
combinations of $\Rey{} \in \{50, 100, 200, 400\}$ and $\kappa \in
\{0.15, 0.22, 0.3\}$. The particle is allowed to rotate in general,
unless otherwise specified.

\begin{figure*}
        \centering
        \includegraphics{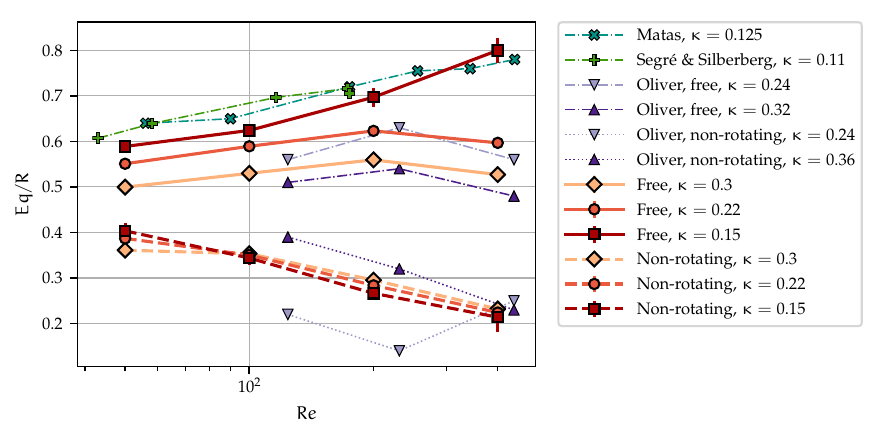}
        \caption{ Experimental data for different $\kappa$ and \Rey{},
          compared to our results. Matas data is
          from~\cite{Matas2004a}, Segr\'e and Silberberg data is
          from~\cite{Segre1962}, Oliver data is
          from~\cite{Oliver1962}.  Data from Oliver corresponds to
          only two particles per experiment.  Solid lines represent
          our results for freely rotating particles, dashed lines -- our non-rotating
          ones. Dash-dotted are experimental rotating, dotted -- experimental non-rotating }
        \label{fig:inertial_rigid_tube_summary}
\end{figure*}

In accordance with the previous numerical and experimental
studies~\cite{DiCarlo2009,Liu2015}, we observe that the bigger
particles equilibrate closer to the channel axis
(\Cref{fig:inertial_rigid_tube_equilibria}\sublabel{a}), and that the
focusing position shifts outwards when \Rey{} is increased
(\Cref{fig:inertial_rigid_tube_equilibria}\sublabel{b}), with the
exception of higher \Rey{} at large $\kappa$.  When the Reynolds
number increases beyond 200, large particles with $\kappa \in \{0.22,
0.3\}$ equilibrate closer to the channel axis, than at $\Rey{} = 200$.
Oliver et al.~\cite{Oliver1962} report a similar behavior with $\kappa
= 0.24$ and $\kappa = 0.32$ (see also \Cref{fig:inertial_rigid_tube_summary}).

In the absence of particle rotation, the equilibrium position of the
particle is closer to the channel axis than in the rotating case (see
\Cref{fig:inertial_rigid_tube_equilibria}).  This result agrees with
previous numerical findings~\cite{Feng1994,Prohm2014}.  However, we
find a much more pronounced change compared to the rotating case than
previously reported.  We conclude from this result that the force due
to rotation of the particle is larger than reported
before~\cite{DiCarlo2009,Amini2014a}.  As also depicted in the \SM~\cite{SM},
the lift forces for the rotation-free particles are negatively shifted
for almost all the cases and all the different particle positions
$y/R$.  That results in the fact that, for example, the small particle
($\kappa = 0.15$) at $\Rey{} = 200$ ($\Rey_p = 4.5$) changes its equilibrium position
from the default $0.7 R$ to $0.28 R$.

The only experimental study found by the authors with non-rotating
particles~\cite{Oliver1962} agrees well with our results.  Particles
during that experiment were prevented from rotation using magnetic
field.  Observed difference in the focusing positions can hardly be
quantified due to the very small size of the dataset, but the major
trends are indisputable.  \cref{fig:inertial_rigid_tube_summary}
compares available experimental data at $\kappa$ similar to ours with
our results.

\begin{figure*}
        \centering
        \includegraphics[width=0.99\textwidth]{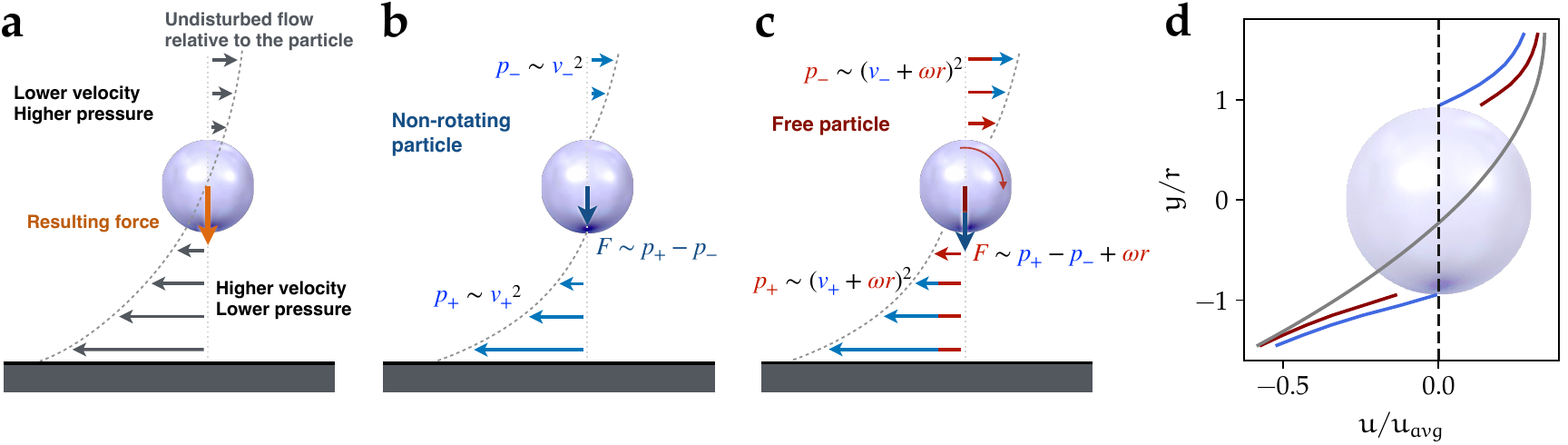}
        \caption{The lift due to the non-uniform flow shear.
          \sublabel{a}: common explanations only takes into account
          the undisturbed flow.  However, disturbed flow over the
          non-rotating particle (\sublabel{b}, in blue) or freely
          rotating one (\sublabel{c}, blue and red) is different,
          being faster close to the surface for the case of the free
          rotation.  This leads to extra lift force attributed to the
          spinning (in red).  \sublabel{d}: numerical experiment shows
          a similar flow velocity profile relative the the
          particle. Simulation parameters: $\Rey{} = 100$, $\kappa =
          0.22$, $y/R = 0.4$ ($\Rey_p = 4.84$). The curves represent averaged
          stream-wise flow velocity along a line through the sphere
          center parallel to $Oy$, blue is non-rotating particle case,
          red is freely rotating one, black is undisturbed flow
          profile far away from the particle. }
        \label{fig:pipe_sketches}
\end{figure*}

To explain the rotation significance, we have to recall the current understanding of the migration mechanism.
According to numerous works~\cite{Feng1994,martel2014,Zhang2016}, the lateral force acting on a particle can be separated into two components: the wall force $F_l^{wall}$, which always pushes the particle away from the wall, and the force due to the non-uniform flow shear rate $F_l^{shear}$.
While the wall repulsion is observed and studied in various problems with sphere translating parallel to the wall~\cite{ZENG2005,Zeng2009,LEE2010}, the shear-induced lift is not understood well.
Its classical explanation attributed to Feng~\cite{Feng1994} is sketched on the \Cref{fig:pipe_sketches}\sublabel{a}.
Looking at the \textit{undisturbed} flow profile relative to the moving particles, one can notice the higher and lower speed regions (shown respectively on the bottom and top of the sketch).
We consider two streamlines that are on each side of the particle but close to each other upstream, where they share the same velocity and pressure.
Thus, applying the Bernoulli principle along each streamline, the difference in velocity on each side of the particle yields a larger pressure above the particle than below it~\cite{Feng1994}.
This results in a net force directed towards the wall.

However, the finite-sized particle always disturbs the flow around itself.
The non-rotating particle slows down the relative flow both above and below (\Cref{fig:pipe_sketches}\sublabel{b}), while the rotating one (\Cref{fig:pipe_sketches}\sublabel{c}) perturbs the background flow less significantly.
Note that the actual data from our simulation in \Cref{fig:pipe_sketches}\sublabel{d} corresponds well with this statement.
This leads to the fact that in the case of rotation the velocity magnitudes both below and above the particle are higher, which, assuming Bernoulli-like quadratic pressure dependence on velocity, results in higher pressure difference across the sphere.
Simple calculations lead to $F_l^{shear}(\omega) \sim \Delta p \sim \Delta p_0 + C \omega$, where $p_0$ is the pressure difference in case of absence of rotation, $\omega$ is the sphere angular velocity and $C$ is a constant.

\begin{figure}
        \centering
        \includegraphics{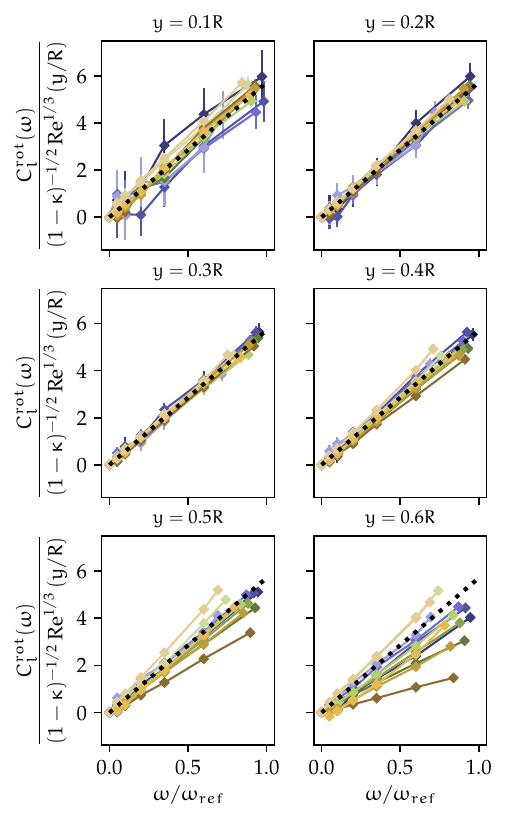}
        \caption{ Scaled rotation-induced lift coefficient for all the
          studied \Rey{} and $\kappa$ is proportional to the rotation
          velocity $\omega$.  Linear dependence remains even when
          particles are close to the wall, $y > 0$, however the slope
          changes in that case.  Dashed line is the empirical fit for
          $y < 0.4 R$. }
        \label{fig:rotation_vs_omega}
\end{figure}

This hypothesis predicts rotation to strengthen the lift towards the
wall, as we observe.  It also implies linear dependence between the
rotation-induced lift force $F_l^{rot}(\omega) = F_l(\omega) - F_l(0)$
and the angular velocity $\omega$.  This behavior is indeed observed,
as depicted in \Cref{fig:rotation_vs_omega}.  The plot shows scaled
difference in the lift coefficient $C^{rot}_l (\omega) = C_l(\omega) -
C_l(0)$ against the rate of rotation $\omega / \omega_{ref}$.  Here
$\omega_{ref} = \dot \gamma / 2$ is the reference angular velocity
(observed when for a sphere rotating in the simple Stokes shear flow)
and $\dot \gamma$ is the shear rate at the sphere center.  We observe
an excellent collapse of all the force-rotation lines for different
\Rey{} and $\kappa$ when $y < R/2$, suggesting a universal scaling
law for the rotation-induced force contribution far from the wall.

In that case, we give the following empirical expression for the
rotation-induced lift coefficient:
\begin{equation}
C_l^{rot}(\omega) = K \frac{\omega}{\omega_{ref}}
(1-\kappa)^{1/2} \Rey^{1/3} (y/R),
\end{equation}
where linear least square fit for $y < R/2$ leads to $K \approx
5.74$.  Moreover, it suggests that the rotation-induced part of the
lift has the same order of magnitude as the pure $F_l^{shear} (0)$,
since in case of the freely-rotating particles its angular velocity is
reasonably close to the reference $\omega_{ref}$.  More detailed force
profiles available in \SM{}~\cite{SM} support that idea.

Finally, the focusing positions of the non-rotating particles show
different behavior with respect to \Rey{} and $\kappa$.  From
\Cref{fig:inertial_rigid_tube_equilibria}\sublabel{a} we see that
bigger non-rotating particles no longer move closer to the pipe axis.
Instead, the focusing position stays roughly constant, only changing
noticeably for the highest $\Rey{} = 400$. More
interestingly, increasing $\Rey{}$ moves the focusing positions further
away from the wall, which is the opposite of the freely rotating
particles (see \Cref{fig:inertial_rigid_tube_equilibria}\sublabel{b}).
The exact reasons of such behavior remain an open question and require
further studies.

\subsection{Square duct}

\begin{figure*}
  \centering
  \includegraphics[width=\linewidth]{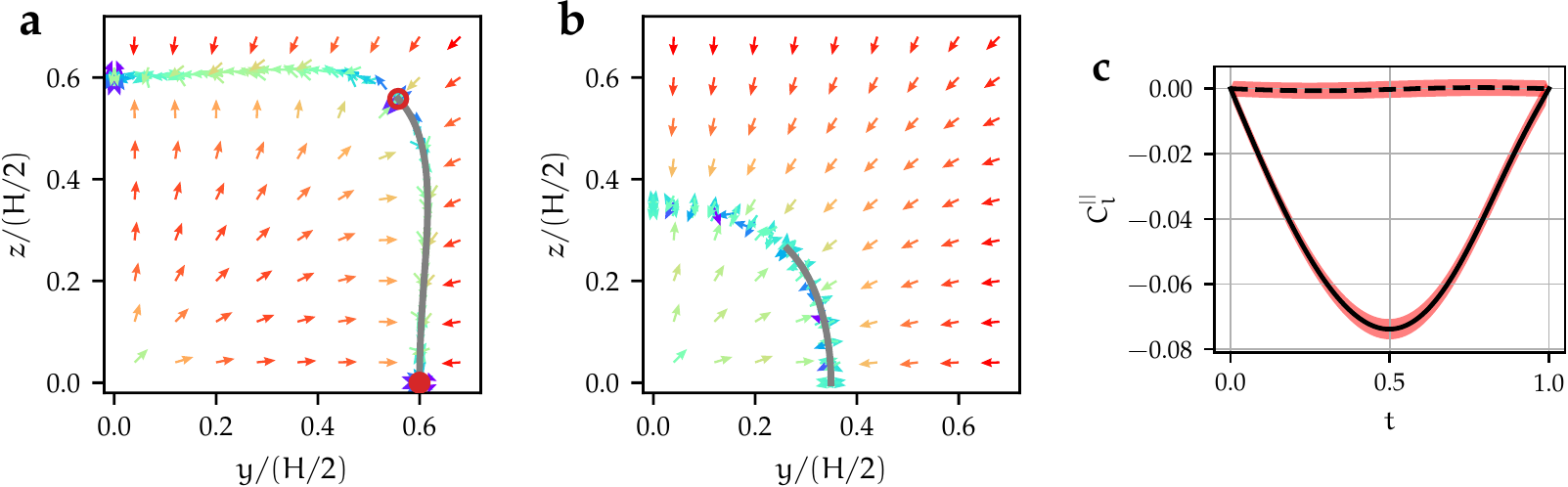}
    \caption{\sublabel{a}, \sublabel{b}: map of the lateral forces
      acting on a sphere, respectively rotating and non-rotating, in a
      square channel at $\Rey{} = 100$, $\kappa = 0.22$.  Each arrow
      corresponds to one simulation and shows the direction of the
      force, while color marks the magnitude (red is higher).  Open
      circle is the unstable equilibrium position, full -- stable,
      gray line is the streamline connecting them (separatrix).
      \sublabel{c}: lift force along the separatrix.  $Ox$ axis is in
      parametric units such that the separatrix manifold is $S =
      \left\{ (x(t), y(t)) \right\}$ with $t = 0$ is the edge
      equilibrium, $t = 1$ is the corner one. }
        \label{fig:inertial_rigid_square}
\end{figure*}

For the case of the square cross-section we only vary $\Rey{} \in \{50, 100, 200\}$, while keeping $\kappa = 0.22$ to reduce the co mputational time.
This corresponds to particle Reynolds numbers $\Rey_p \in \{2.42, 4.84,  9.68\}$.
For the sake of brevity we hereafter focus on $\Rey{} = 100$,
noting that the general findings for the other Reynolds numbers are
similar according to our results.
\Cref{fig:inertial_rigid_square}\sublabel{a} shows the map of lift
forces acting on a freely rotating particle at different lateral
positions.  Due to the symmetry of the square channel, we only perform
simulations for one half of the cross-section quadrant ($y>0$, $z>0$,
$z \leqslant y$).  For illustrative purposes, we plot forces for the
full quadrant by mirroring them along the diagonal $y = z$.

Analysis of the lateral force vector field reveals two different types
of equilibrium positions consistent with the previous experimental and
numerical studies\cite{Choi2011,Miura2014,Nakagawa2015,Liu2015}: the
corner position is an unstable saddle point, as some force vectors
point out of it, and the mid-edge position is stable.  The streamline
starting in the unstable equilibrium and finishing in the stable one
(\textit{separatrix}) divides the lateral positions into two zones: if
the particle starts close to the pipe axis, it will initially migrate
towards the wall, while the particle starting between the separatrix
and the wall will move to the center.  While on the separating
streamline, the particle will continue moving along it until it
reaches the stable edge equilibrium.  As a characteristic of the
secondary migration speed, in
\Cref{fig:inertial_rigid_square}\sublabel{c} we plot the force acting
on the particle along the separatrix curve.  The parametrization of
the separatrix $\vec S(t) = (x(t), y(t))$ is such
that $\vec S(0)$ is the stable edge-center equilibrium, and $\vec
S(1)$ is the unstable one.  The force along the separatrix can
therefore be calculated as follows:
\begin{equation}
        F^{||}_l(t) = \vec F_l(\vec S(t)) \cdot \lb \frac{d \vec
          S}{dt}(t) \left\lvert \frac{d \vec S}{dt}(t)
        \right\rvert^{-1} \rb.
\end{equation}
Here $\cdot$ means the dot product.

We do not compare fast and slow migration as it has been done
previously~\cite{Nakagawa2015}, but instead we draw our attention to
the effect of inhibited rotation.
\Cref{fig:inertial_rigid_square}\sublabel{b} shows the lateral force
map for the \textit{non-rotating} particle.  In this case we can not
anymore clearly identify specific equilibrium points: instead we
observe a one-dimensional equilibrium manifold.  All the points on
that line result in zero force and migration velocity, as seen in
\Cref{fig:inertial_rigid_square}\sublabel{c}.  Hence we conclude that
the particles prevented from rotation undergo \textit{no secondary
  migration}.  To the best of our knowledge that has not been reported
neither experimentally nor numerically before.

\begin{figure}
  \centering
  \includegraphics{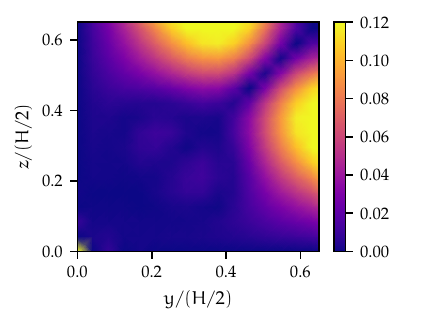}
  \caption{Heat map corresponds to the dot product of unit vectors in
    the direction of the rotation-induced force $F_l^{rot}$ and angular velocity $\omega$,
    representing their alignment.}
  \label{fig:inertial_force_vs_rot}
\end{figure}

One can consequently think that rotation is responsible for the
particle movement along the separatrix towards the stable equilibrium
(see, e.g.~\cite{Zhou2013}).  To assess this hypothesis, we take two
steps.  As the first step, we show that the direction of the rotation-induced
force is highly correlated with the vector, normal to the particle's angular
velocity.  We subtract two force-fields (with and without rotation)
and compute the angle $\phi$ between the difference $\vec{F}^{diff}$
and the angular velocity vector of the freely rotating particle.
\Cref{fig:inertial_force_vs_rot} shows that the angle between the two
vectors perpendicular within less than
\SI{10}{\degree}, with more pronounced discrepancy closer to the wall.
This observation suggests that small changes in rotation induce force
roughly \textit{orthogonal} to that change, and that only the rotation
\textit{orthogonal} to the separatrix line can cause the secondary
migration.

\begin{figure}
        \centering
        \includegraphics{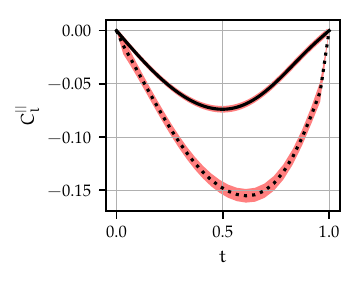}
        \caption{Force along the separatrix driving a particle from
          the corner to the edge equilibrium (solid line) increases if
          particle is prevented from rotating normally to the
          separatrix curve (dashed).}
        \label{fig:inertial_no_normal_rot}
\end{figure}

In the second step, we check our supposition by carrying out
simulations with modified rotation of the particles on the separatrix.
In these simulations, we allow the particles to freely rotate
tangentially to the separatrix line, while the normal component of the
rotation is modified and set to zero.  In agreement with our previous
conclusion, the resulting force only differs tangentially to the
separatrix, while the normal component (that would have changed the
location of the separatrix) stays zero.  However, contrary to our
initial hypothesis, the force of the secondary migration is increased,
see \Cref{fig:inertial_no_normal_rot}.  Indeed, such direction of the
rotation-induced force agrees with previous
works~\cite{Oliver1962,Feng1994}, and with our results for the
circular pipe.  This result means that the secondary migration is not
solely attributed to the rotation, but is instead governed by the
interplay of all the forces including wall force, shear- and
rotation-induced components.  The exact physical picture could be the
focus of later studies.

\begin{figure}
  \centering
  \includegraphics{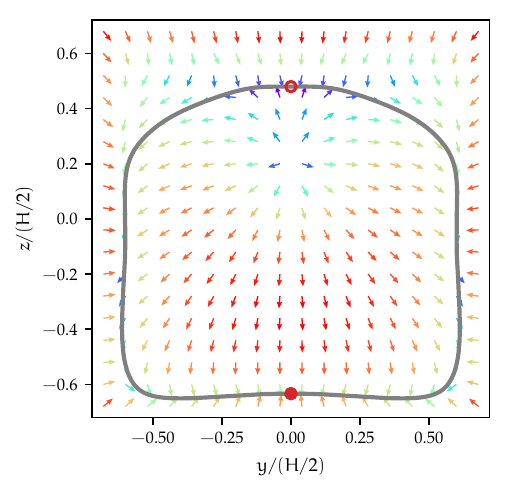}
  \caption{ When extra torque is applied to a particle the
    equilibrium positions change, with only a single stable edge
    equilibrium position apparent.  Arrows are map of the
    lateral forces acting on a sphere in a square channel at
    $\Rey{} = 100$, $\kappa = 0.22$ ($\Rey_p = 4.84$) \textbf{with applied external
      torque along $Oy$}: $T^{ext}_y = 0.75 \rho U_{avg}^2
    (2r)^5 H^{-2}$.  Each arrow corresponds to one simulation
    and shows the direction of the force, while color marks the
    magnitude (red is higher).  The gray line is the equilibrium
    manifold. }
  \label{fig:inertial_extra_torque}
\end{figure}

Finally, the fact that rotation produces close to orthogonal force
allows us to consider modification of the equilibrium positions by
manipulating rotation.  A similar approach was presented
in~\cite{Prohm2014}, however only in 2D.  We apply external torque
$\vec T^{ext}$ in the positive direction of the $Oy$ axis which
results in shifting the lateral force-field and breaking the symmetry
(see \Cref{fig:inertial_extra_torque}).  The extra rotation overcomes
the intrinsic force of the secondary migration and results in
elimination of the stable equilibria at $y = 0$.  The corner positions
disappear as well, and the top one of the remaining two equilibria
becomes unstable.  Therefore all the particles experiencing external
torque (high enough to overcome the natural inertial forces), will now
gather at the unique lateral position.

\section{Conclusions}
In this work we employed the \DPD method to numerically examine
lateral inertial migration of rigid spherical particles and focused
our attention to the role of particle rotation in the migration.  Our
results for circular pipe show that rotation significantly changes
particle equilibrium position despite the common thinking that its
contribution is negligible.  Moreover, we observed that non-rotating
particles focus closer to the pipe axis with increased Reynolds, which
is the opposite to the freely rotating ones.  We also proposed
phenomenological scaling of the rotation-induced lift coefficient with
\Rey{}~and~$\kappa$.

Additionally, we studied channels with a square cross-section.  We
found that particles with restricted rotation undergo no secondary
migration, remaining on the equilibrium annulus.  Moreover, we proved
wrong a seemingly logical hypothesis that rotation solely moves
particles during the second migration stage: instead this stage seems
to be governed by the interplay of all the forces in the system, with
rotation actually slowing the migration along the annulus.  Finally,
we proposed a simple way to modify the focusing in the square duct,
removing all but just one equilibrium position.  Our findings
suggest novel inertial microfluidic designs that exploit the  rotation of the
focusing particles.

\section*{Acknowledgments}

We acknowledge the computational resources provided by the Swiss National Supercomputing Center (CSCS) under project IDs ch7, and by the FASRC Cannon cluster, supported by the FAS Division of Science Research Computing Group at Harvard University.

\bibliographystyle{unsrt}
\bibliography{refs}
\end{document}


\title{Supplementary material: \\
Inertial Focusing of Spherical Particles: \\ The Effects of Rotational Motion}
\date{\today}

\maketitle

\section{Numerical method}
\label{app:sec:numerics}

\ymr implements classical \DPD method, which yields fluctuating
hydrodynamics~\cite{Espanol1995,Groot1997}.  The evolution of the
system is governed by the pairwise particle forces with enforcing of
the no-slip and no-through boundary conditions where applicable.

\subsection{Dissipative particle dynamics}

We employ \DPD, a particle mesoscale method introduced by Hoogerbrugge
\cite{Hoogerbrugge1992} and revisited by Groot and
Warren~\cite{Groot1997} and Espanol~\cite{Espanol1995}.  The DPD fluid
is described in terms of a set of identical particles in the 3D space.
Each particle is characterized by its mass $m$, position $\vec{r}$ and
velocity $\vec{v}$.  Particles evolve in time according to the
Newton's law of motion:
\begin{equation}
\begin{aligned}
\frac{d \vec{r}}{dt} &= \vec{v}, \\
\frac{d \vec{v}}{dt} &= \frac {1}{m} \vec{F},
\end{aligned}
\end{equation}
where $\vec{F}$ is the force exerted on the particle and $t$ is time.
The force magnitude is typically bound for any $r$, and vanish after a
cutoff radius $r_c$.  The particles interact through central forces,
which implies, by the Newton's third law, conservation of linear and
angular momenta.  The \DPD forces acting on the particle indexed by
$i$ are written as
\begin{equation}
\vec{F}_i = \sum\limits_{j} \left( \vec{F}_{ij}^C + \vec{F}_{ij}^D + \vec{F}_{ij}^R \right),
\end{equation}
where the force has been split into three parts: conservative,
dissipative and random.  The conservative term acts as purely
repulsive force and reads
\begin{equation}
\vec{F}_{ij}^C = a w(r_{ij}) \vec{e}_{ij},
\end{equation}
where $r_{ij} = |\vec{r}_{ij}|$, $\vec{r}_{ij} = \vec{r}_i -
\vec{r}_j$, $\vec{e}_{ij} = \vec{r}_{ij} / r_{ij}$ and
\begin{equation}
w(r) = \begin{cases}
1 - r/r_c, & \text{if } r < r_c, \\
0,         & \text{otherwise}.
\end{cases}
\end{equation}
The dissipative and random terms are given by
\begin{equation}
\begin{aligned}
\vec{F}_{ij}^D &= - \gamma \left( \vec{v}_{ij} \cdot \vec{e}_{ij} \right) w_D(r_{ij}) \vec{e}_{ij},\\
\vec{F}_{ij}^R &= \sigma \xi_{ij} w_R(r_{ij}) \vec{e}_{ij}.
\end{aligned}
\end{equation}
The random variable $\xi_{ij}$ is independent Gaussian noise
satisfying $\langle \xi_{ij}(t) \xi_{lm}(t') \rangle = \delta(t-t')
\left( \delta_{il} \delta_{jm} + \delta_{im} \delta_{jl} \right)$,
$\xi_{ij} = \xi_{ji}$ and $\langle \xi_{ij} \rangle = 0$.
The parameters $\gamma$ and $\sigma$ are linked through the
fluctuation-dissipation relation $w_D = w_R^2$ and $\sigma^2 = 2\gamma
\kbt$~\cite{Espanol1995}.  We usually choose the dissipative kernel as
$w_R(r) = w^k(r)$, $k \in (0, 1)$~\cite{Fan2006}.

Rigid objects are composed of an analytical surface and a set of DPD particles that have fixed positions relative to the rigid object's  frame of reference.
Rigid objects are thus fully described by their center of mass, orientation, described by a quaternion, linear velocity and angular velocity.
These quantities are advanced in time using a velocity-Verlet integration scheme adapted to quaternions~\cite{rozmanov2010robust}, using the moment of inertia computed analytically from the object's shape.

In order to push the flow along a channel we apply the pressure
gradient $\nabla p = \vec f \rho$, where $\vec f$ is the body force,
i.e. the force applied to every liquid \DPD particle.

\subsection{Boundary conditions}

The geometrical boundaries of the simulation, or walls, are represented as the zero-isosurface of a Signed Distance Function (SDF).
A layer of frozen particles with thickness $r_c$ is located inside the boundary.
These particles have the same radial distribution function as the fluid particles, and interact with the latter with the same \DPD forces.
This ensures the no-slip condition as well as negligible density variation in proximity to the wall.
In addition, particles are bounced-back from the surface, ensuring no-through condition on the wall surface~\cite{Revenga1998}.

The fluid-structure interactions describing rigid objects boundary conditions are similar to those prescribed at the walls.
The object surface is represented as an analytical SDF representing a sphere.
The surface impenetrability is enforced by bouncing-back solvent particles from that surface.
The momentum change of the bounced particles is transferred to the object force and torque, to ensure total linear and angular momentum conservation.

\section{Convergence of the lift coefficient with domain length}

In this study, periodic boundary conditions are used along the pipe direction.
In this case, periodic images may disturb the system.
Thus, we select the length of the domain along the periodic direction, $L$, to be large enough to avoid such periodic effects.
We vary this length up to $L=20$ in particle diameters units, in a circular pipe at $\Rey=50$, $\kappa=0.15$, and DPD parameters $a=160$, $s=0.5$, $\gamma=40$, $\rho=8$, $m=1$, $k_BT=3$ and a particle radius $r=5 r_c$, where $r_c=1$ is the DPD cutoff radius.
An external force of magnitude $0.595$ is exerted on each DPD particle in the flow direction, to mimic the pressure gradient.
The particle's position is constrained to a radial position $0.4 R$.
The system is evolved for a total time $T=600 H / U_\text{avg}$, and the lift coefficient $C_l$ is computed from the average force over the three last quarters of the simulation time, starting well after equilibration of the system.

\begin{figure}
  \centering
  \includegraphics{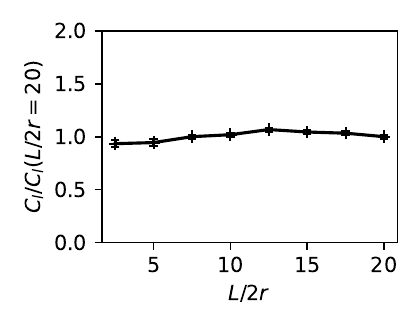}
  \caption{Lift coefficient $C_l$, normalized by the value obtained at $L/2r = 20$, against the periodic domain length $L$.}
  \label{fig:lift:convergence}
\end{figure}

\Cref{fig:lift:convergence} shows that the value of the lift coefficient converges as $L$ increases, and we select $L/2r = 15$ in the rest of this study.
This value is a good compromise between accuracy and computational cost.

\section{Method verification}

Here we discuss the applicability of the presented DPD method to the
inertial migration problems.  A recent work by Huang et
al.~\cite{Huang2018} brought up the main issues of the DPD regarding
particle focusing in the plane Poiseuille flow: (i) high particle
diffusivity related to the low solvent Schmidt number \Sc; (ii) high
solvent compressibility leading to excessive Mach number \Ma; (iii)
solvent shear-thinning, which breaks hydrodynamic DPD behavior.
Indeed, we encountered similar problems and approached them, in
general, by increasing spatial or temporal resolution of the
simulations made possible by the high throughput of \ymr.

\begin{figure}
        \centering
        \subcaptionbox{Viscosity of the DPD solvent depending on the pipe Reynolds number at different values of $\gamma$: 2 (squares), 10 (circles), 40 (diamonds). The other DPD parameters are as follows: $\rho = 8, r_c = 1, a = 160, \kbt = 3.0, s = 0.5, \delta t = 5\times10^{-4}$. \label{fig:dpd_thinning}}
        {
                \includegraphics{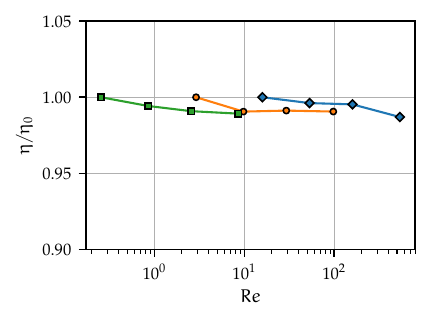}
        }\hspace{0.5cm}
        \subcaptionbox{Particle lift coefficients in a pipe ($\Rey = 50$, $\kappa = 0.15$) depending on the DPD temperature.
                Inset: increased $\kbt$ expectedly increases standard deviation of the forces.	\label{fig:schmidt}}
        {
                \includegraphics{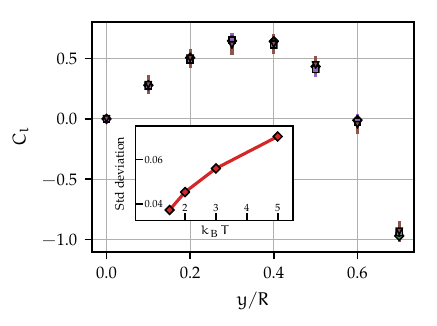}
        }\hspace{0.5cm}
        \subcaptionbox{Particle lift coefficients in a pipe ($\Rey = 50$, $\kappa = 0.15$) depending on the Mach number. Triangles correspond to $\Ma=0.36$, squares -- $0.14$, circles -- $0.1$ and diamonds -- $0.07$. \label{fig:mach}}
        {
                \includegraphics{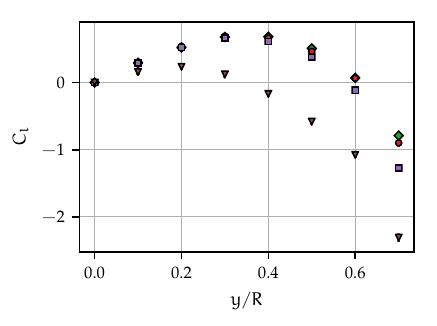}
        }
        \caption{}
        \label{fig:convergence}
\end{figure}

In order to assess how the particle diffusivity, or Brownian motion, affects the measured lateral inertial forces, we perform simulations at different $\kbt$, which governs $Sc \sim \lb \kbt\rb^{-1}$~\cite{Groot1997}.
\Cref{fig:schmidt} shows the lift forces for $\Rey = 50, \kappa = 0.15$ and their average standard deviation depending on the temperature.
The other DPD parameters are fixed and will be listed later in this section.
We see clear independence of the mean force with respect to $\kbt$ and thus \Sc, with only the variance increasing as \Sc is decreasing.
Such observations agree with Huang et al.~\cite{Huang2018}.
Pursuing higher \Sc in order to reduce the variance, we use the kernel exponent $s = 0.5$ following Fan et al.~\cite{Fan2006}.

The Mach number $\Ma = 2 u_{avg} r / (H c_s)$, on the other hand, plays a crucial role in the simulation results.
The effect of changing \Ma by varying repulsion parameter $a$ is shown in \cref{fig:mach}.
Conceivably, the results are converging with decreased \Ma.
The lowest $\Ma = 0.07$ corresponds to $\Ma^H = 0.45$ in notation of Huang et al.~\cite{Huang2018}, which is lower than their threshold value of $0.8$.
We believe that not low enough \Ma may be the primary reason for Huang's mediocre experimental agreement, especially in the trend of equilibrium moving towards the channels axes with increased \Rey.
We also note that further increase in \Ma makes the simulations more and more expensive, as high repulsion $a$ requires lower time-step and higher $\kbt$ to prevent ``freezing''~\cite{Fedosov2008}.
A known relation (see, e.g,~\cite{Groot1997}) for the DPD speed of sound $c_s$ reads
\begin{equation}
c_s^2 = \kbt + 2\alpha a \rho,
\end{equation}
with $\alpha \approx 0.101$.
So to increase $c_s$ we always use high DPD density $\rho = 8$ and scale $a$ as needed.

Finally, we select the DPD parameters such that shear-thinning is negligible: see \cref{fig:dpd_thinning}.
We observe that at high $\rho=8$ the shear-thinning in much less pronounced at higher $a$ than in Huang et al.~\cite{Huang2018}.
So our final values look as follows: $m=1$, $\rho = 8$, $r_c = 1$, $a = 160$, $\kbt = 3.0$, $s = 0.5$, $\gamma \in [1, 100]$, $\delta t \in [10^{-5}, 10^{-4}]$.
Viscosity in this setup is solely governed by the dissipative parameter $\gamma$, which we adjust according to \Rey.
Finally, the resolution of the simulations is set by the particle size $r = 5r_c$, where $r_c=1$ is the cutoff radius of the DPD interactions.

\begin{figure}
        \centering
        \subcaptionbox{$\Rey = 50$, $\kappa = 0.15$. Blue circles are data extracted from the Supplementary Material of Liu\cite{Liu2015}, green diamonds are present simulations. \label{fig:inertial_val_circle_lift}}
        {
                \includegraphics{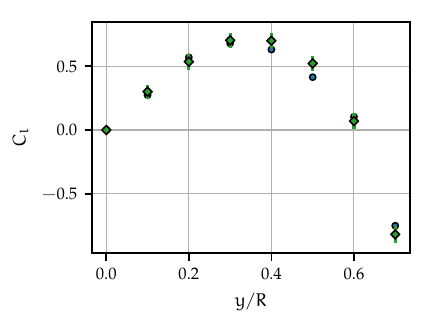}
        }
        \hspace{1cm}
        \subcaptionbox{$\Rey = 40$, $\kappa = 0.22$. Blue circles are data from Nakagawa\cite{Nakagawa2015}, orange squares from Di Carlo\cite{DiCarlo2009}, green diamonds are present simulations. \label{fig:inertial_val_square_lift}}
        {
                \includegraphics{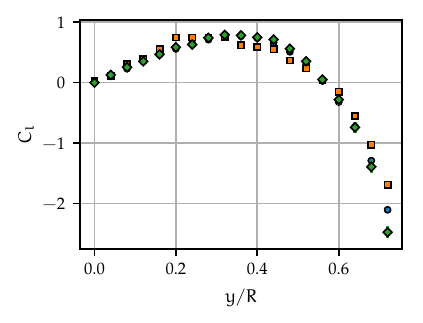}
        }
        \caption{Sphere lift coefficients for the circular \subref{fig:inertial_val_circle_lift} and square \subref{fig:inertial_val_square_lift} pipes. $z$ position of the particle is on the channel axis and $y$ position varies.}
\end{figure}

Now we compare our method against several numerical and experimental
results.  First we compute the lift coefficients for the freely
rotating rigid particles translating in the circular pipe, see
\Cref{fig:inertial_val_circle_lift}.
We observe a good agreement with the numerical results reported in the Supplementary Material of Liu et al.~\cite{Liu2015}.
Next we examine the lift in the square channel flow, see \Cref{fig:inertial_val_square_lift}.
We observe a good overall agreement with \cite{Nakagawa2015}, with the only significant
difference (${\sim}15\%$) when the particle is closer than $0.03 H$ to
the wall.  However, two numerical calculations we compare
against\cite{Nakagawa2015,DiCarlo2009}, do not agree with each other
in that region, which suggests that the data for the last point may be
unreliable for all the methods.  In our following simulations we keep
the wall distance to be at least $0.04 H$.

\begin{figure}
        \centering
        \begin{subfigure}[t]{0.32\textwidth}
                \centering
                \includegraphics[width=1.0\textwidth]{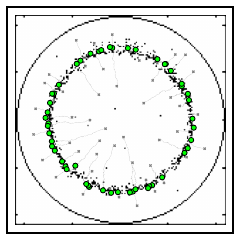}
                \caption{$\Rey = 58$, $\kappa = 0.083$, circular cross-section, experiments from \cite{Choi2010}}
        \end{subfigure}%
        \hfill
        \begin{subfigure}[t]{0.32\textwidth}
                \centering
                \includegraphics[width=1.0\textwidth]{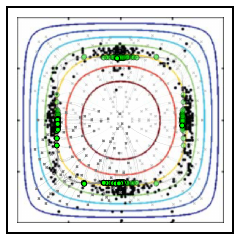}
                \caption{$\Rey = 57$, $\kappa = 0.15$, square cross-section, experiments from \cite{Choi2011}}
        \end{subfigure}%
        \hfill
        \begin{subfigure}[t]{0.32\textwidth}
                \centering
                \includegraphics[width=1.0\textwidth]{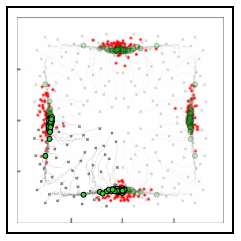}
                \caption{$\Rey = 144$, $\kappa = 0.108$, square cross-section, experiments from \cite{Miura2014}}
        \end{subfigure}
        \caption{Experimental focusing positions with overlays of
          simulated trajectories of the free rigid particles projected
          on the cross-section plane. Crosses mark the initial
          positions, full circles mark the final positions after the
          particle has traveled $1000 R$ in the streamwise
          direction. Experimental data in black or red, our numerical
          data in green. Opaque symbols represent actual simulations,
          while transparent ones are symmetric copies drawn for
          clarity. }
        \label{fig:val_channel_migration}
\end{figure}

Finally we compare the simulation results for the unconstrained
migration of rigid particles in the circular and square channels, see
\Cref{fig:val_channel_migration}.  The final positions here are
obtained without restricting the particle motion at all, allowing them
to settle laterally into the equilibrium after traveling $L=500H$
downstream.  The obtained focusing positions overlap well with the
experimental results of Choi\cite{Choi2010} and Miura\cite{Miura2014}
and show slight discrepancy with the data of Choi\cite{Choi2011}.  We
suppose that the latter mismatch may be attributed to the fact that
the channel in Choi\cite{Choi2011} has a slightly non-square
cross-section as described in the Supporting Information, with a
little different width and length and rounded corners.  This possibly
leads to a noticeable shift in the equilibrium positions.

\section{Detailed data}
\begin{figure*}
        \centering
        \includegraphics{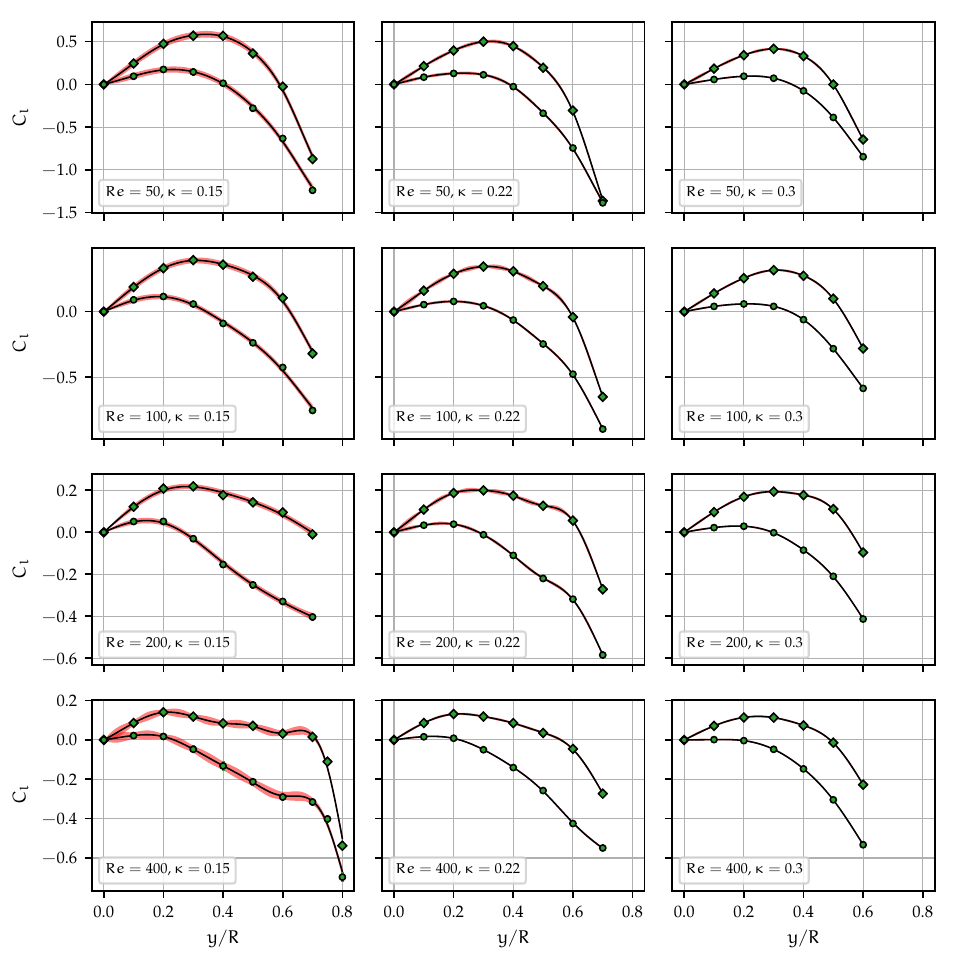}
        \caption{Summary of the lift coefficients for different
          channel Reynolds numbers \Rey and particle-to-channel size
          ratios $\kappa$. Diamonds are freely rotating particles,
          circles -- prevented from rotating. }
        \label{fig:inertial_rigid_tube_all}
\end{figure*}

Here we present the lifting force plots for all the studied cases in
circular pipe (\Cref{fig:inertial_rigid_tube_all}).

\bibliographystyle{unsrt}
\bibliography{refs}